\documentclass[pre,amssymb,amsmath,floats,twocolumn,showpacs,superscriptaddress]{revtex4}

\usepackage{epsfig}

\begin{document}

\title{Complex networks created by aggregation}

\author{M. J. Alava}
\email{Mikko.Alava@hut.fi}
\affiliation{Helsinki University of Technology, Laboratory of Physics, HUT-02105 Finland}

\author{S. N. Dorogovtsev}
\email{sdo@fyslab.hut.fi, sdorogov@fis.ua.pt}
\affiliation{Helsinki University of Technology, Laboratory of Physics, HUT-02105 Finland}
\affiliation{Departamento de F{\'\i}sica da Universidade de Aveiro, 3810-193 Aveiro, Portugal}
\affiliation{A. F. Ioffe Physico-Technical Institute, 194021
  St. Petersburg, Russia}

\date{}

\begin{abstract}
We study aggregation as a mechanism for the creation of complex networks. 
In this evolution process vertices merge together, which increases a number of highly connected hubs.   
We study a range of complex network architectures produced by the aggregation. 
Fat-tailed (in particular, scale-free) distributions of connections are obtained both for networks with a finite number of vertices and growing networks. 
We observe a strong variation of a network structure with growing density of connections   
and find the phase transition of the condensation of edges. 
Finally, we demonstrate the importance of structural correlations in these networks. 
\end{abstract}


\maketitle

\section{Introduction}\label{s-introduction}

Fat-tailed distributions of connections characterize the complex architectures of many real-world networks \cite{ab02,dmbook03,n03,pvbook04}. Several mechanisms may accounted for this form of degree distributions of networks. (Degree is the total number of connections of a vertex.) The most popular concepts imply self-organization \cite{p76,ba99,baj99}.
 The self-organization mechanism is responsible for fat-tailed distributions in a wide circle of evolving systems (see, e.g., Refs. \cite{y25,s55}), and not only in networks. 

Usually, 
a 
very particular preferential attachment version
of the self-organization mechanism 
is 
 discussed, so that highly connected vertices preferentially attract new connections \cite{p76,ba99,baj99}, but there are other possibilities. In this paper, we consider agglomeration as a competing possibility. 
 It is known that aggregation processes effectively generate power-law distributions (see, e.g., Ref.~\cite{t89} for 
an example). In networks, the analogue is the merging of vertices. By this mechanism, 
vertices accumulate their connections (agglomeration of edges).  
This increases a number of highly connected hubs and so gives a chance to arrive at a fat-tailed degree distribution. 

Evidently, the merging of vertices should take place in cellular networks (merging proteins) as well as in many other real-world networks. For example, in various networks of economic relations, merging and splitting of enterprises are basic elements of the evolution. The same is valid for networks of software components, electronic circuits, networks of relations between social groups, organizations, institutions, and parties, networks of subjects, networks of notions, etc.  
Simple evolving networks with merging vertices have quite recently been
simulated, Ref.~\cite{ktms04}, and the generation of fat-tailed degree distribution has been successfully demonstrated. (For a similar process in bipartite graphs, see Ref.~\cite{srtm04}).

In the present paper we provide a comprehensive description of the process of the creation of complex network architectures by the merge of vertices. 
It is impossible to obtain a uniform picture for all networks of this type. 
So, we describe a set of typical types of behaviors by considering a line of basic network models, which can be studied analytically. 
These models may be generalized in a natural way to include clustering and the condensation of clustering. Many other variations are also possible. 


All the models that we study in this paper generate fat-tailed degree distributions. 
We consider both non-equilibrium networks with a fixed number of vertices, and networks, where this number grows. 
We use the mean degree $\overline{k}$ of a network as a relevant parameter. Then the variation of network architectures with $\overline{k}$ is essentially characterized by the $\gamma(\overline{k})$ dependence. So, our main results are presented in the form: exponent $\gamma$ vs. the mean degree $\overline{k}$.  

In most of the networks in this paper, the evolution is due to two parallel processes: (i) the merging of vertices and (ii) random attachment of new vertices. 
However, we also discuss networks, where the second channel of the evolution is splitting (fragmentation) of vertices.The range of scenarios is wide, but in most of them we find a phase with the condensate of edges (in other words, gelation). Above some critical value $\overline{k}_c$ of the mean degree, a finite fraction of edges is attached to a few vertices or to a single vertex. 
This condensation, unlike the situation described in Ref.~\cite{bb01}, occurs in the homogeneous networks. 
In the ``normal phase'', a degree distribution is of a power-law form, $P(k) \propto k^{-\gamma}$. Moreover, we observe that, rather unexpectedly, even in the condensation phase, normal vertices have a scale-free degree distribution. 


A resulting picture may be complicated by the presence of correlations, which is typical for non-equilibrium networks. We demonstrate the importance of degree--degree correlations, in the most succinct of these network models.
The paper is organized as is follows. 
In Sec.~\ref{s-results}, we describe the models and present in detail our results for each of them. The complete final information can be obtained from this section.   
In Sec.~\ref{s-derivations} we present details of our analytical calculations and simulations.

\section{Models and results}\label{s-results}

In this section we describe basic models of networks evolving due to aggregation processes and present our results. For sake of brevity we consider only undirected networks, i.e., networks with undirected edges.

\subsection{Network $O$}\label{ss-o}

This is the simplest model. The evolution starts from a given configuration of vertices and connecting edges. Loops of length one are allowed. 
At each time step (see Fig.~\ref{f1}): 


\begin{figure}
\epsfxsize=50mm
\epsffile{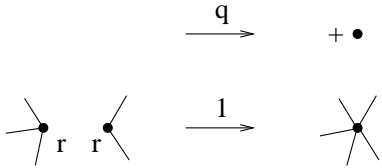}
\caption{
Processes creating network $0$. The labels $r$ indicate that merging vertices are selected at random. ``$+$'' indicates an added vertex. $q \geq 1$ is a relative rate of the addition process. 
}
\label{f1}
\end{figure}


\begin{itemize}

\item[(1)]
$q \geq 1$ new bare vertices are added to the network. 

\item[(2)]
Two randomly chosen vertices merge. 

\end{itemize}

Obviously, the final state of the network is a set of bare vertices plus a single vertex with all the edges (actually, loops of length one) attached. This is what we call the condensate of edges.

\subsection{Network A}\label{ss-a}

The (large) number of vertices of this network, $N$, does not change during the evolution. Initially, there is an arbitrary configuration of $N$ vertices connected by some number of links. At each time step (see Fig.~\ref{f2}): 


\begin{figure}
\epsfxsize=58mm
\epsffile{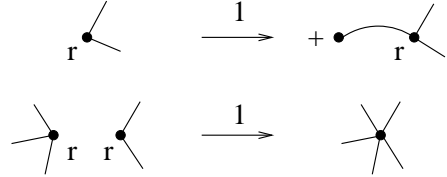}
\caption{
Processes creating network A. 
}
\label{f2}
\end{figure}


\begin{itemize}

\item[(1)]
A new vertex is added to the network. This vertex is attached to a randomly chosen vertex.  

\item[(2)]
Two randomly chosen vertices merge. 

\end{itemize}
So, the result of the merging of vertices of degrees $k'$ and $k''$ is a vertex of degree $k = k'+k''$. The total degree of the network linearly grows, $K(t) = K(t=0) + 2t$, as well as its mean degree. Network becomes more and more dense with time. 

The resulting degree distribution is of a power-law (an asymptotics) form with exponent equal to $3/2$: 

\begin{equation}
P(k) \sim k^{-3/2}
\, .
\label{e1}
\end{equation} 
The main part of the distribution is stationary, but its low-degree part and a cut-off in the high-degree range change with time. This ensure the growth of the mean degree with time. Condensation of edges is absent. 

Note that we assume that this network is sparse. 
In principle, we allow multiple connections and 1-loops. However, we believe that in the sparse network regime, they are not important if we are not interested in the position of the cutoff of the degree distribution.  

This network was simulated earlier \cite{ktms04}. Our analytical results confirm the observations in Ref.~\cite{ktms04}.

\subsection{Network B}\label{ss-b}

This is a growing version of model A. At each time step (see Fig.~\ref{f3}): 


\begin{figure}
\epsfxsize=58mm
\epsffile{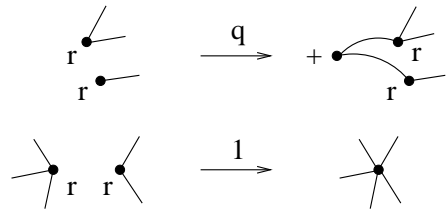}
\caption{
Processes creating network B. Here, $m=2$. 
}
\label{f3}
\end{figure}


\begin{itemize}

\item[(1)]
$q>1$ new vertices are added to the network. Each of these vertices is attached to a randomly chosen vertex by $m$ edges.   

\item[(2)]
Two randomly chosen vertices merge. 

\end{itemize}
The total number of vertices now grows: $N(t)= N(t=0)+(q-1)t$. 
The total degree is $K(t) = K(t=0)+2qmt$. So, the average degree approaches the finite value $\overline{k}=2qm/(q-1)$. 

The condensation of edges is absent in this model.   
Unlike model A, the stationary degree distribution of this growing network is a rapidly decreasing function. If, however, the rate of the growth is low, $q-1 \ll 1$, then the power-law dependence (\ref{e1}) is realized in the range of degrees below a size-independent cutoff, $k \ll m/(q-1)^2$.

\subsection{Network C}\label{ss-c}

In this growing network, at each time step (see Fig.~\ref{f4}): 


\begin{figure}
\epsfxsize=58mm
\epsffile{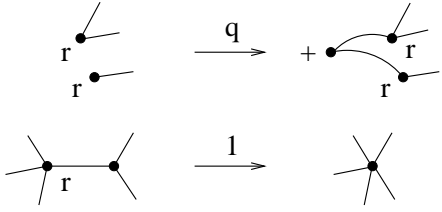}
\caption{
Processes creating network C. $m=2$.  
}
\label{f4}
\end{figure}


\begin{itemize}

\item[(1)]
$q>1$ new vertices are added to the network. Each of these vertices is attached to a randomly chosen vertex by $m$ edges.   

\item[(2)]
Simultaneously, a randomly chosen vertex merges with its randomly chosen neighbor, and the connecting edge disappears.   

\end{itemize}
Rule (2) means the preferential choice: the second vertex in the network is chosen with probability proportional to its degree.  
The number of vertices and the total degree grow as $N(t)= N(t=0)+(q-1)t$ 
and 
$K(t)= K(t=0)+2(qm-1)t$, respectively. So, the mean degree approaches 
the value  

\begin{equation}
\overline{k} \cong 2\frac{qm-1}{q-1} 
>2m
\, .
\label{e2}
\end{equation}  

It turns out that if the network is sufficiently dense, namely if 

\begin{equation}
\overline{k} > \overline{k}_c = 2m(1 + \sqrt{1-1/m})
\, , 
\label{e3}
\end{equation}  
than a finite fraction of edges is condensed on a single vertex (or, maybe, on a few vertices). This takes place when the rate of the grows is low: 

\begin{equation}
q < q_c = 1 + \sqrt{1-1/m}
\, .  
\label{e4}
\end{equation} 


\begin{figure}
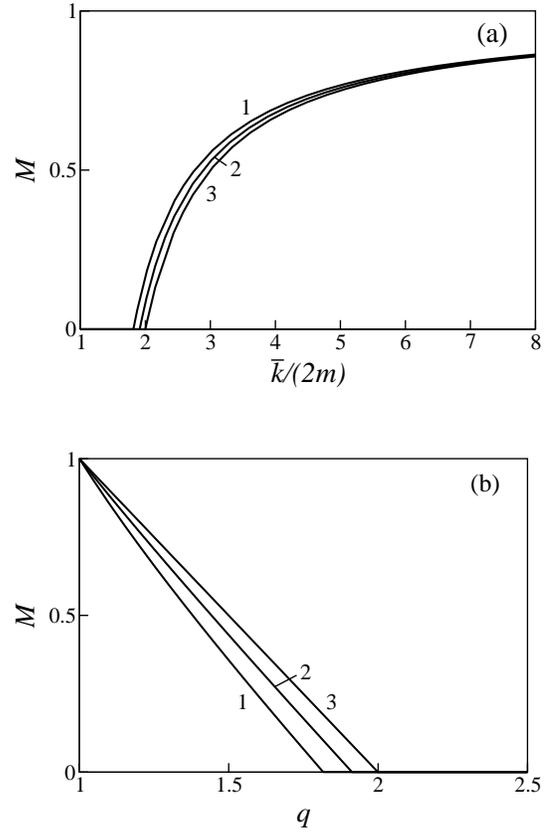

\epsfxsize=70mm
\epsffile{aggregation_fig5a.eps}
\epsfxsize=70mm
\epsffile{aggregation_fig5b.eps}
\caption{
The fraction $M$ of condensed edges in network C as a function of the $\overline{k}/(2m)$ (a) and of the growth rate $q$ (b). Curves 1, 2, and 3 correspond to $m=3,$ 6, and $\infty$, respectively. 
}
\label{f5}
\end{figure}


The fraction of edges in the condensate 

\begin{equation}
M = \frac{\overline{k}^2 - 4\overline{k}m + 4m}{\overline{k}(\overline{k}-2m)} = \frac{2qm-q^2m-1}{qm-1} 
\, .  
\label{e5}
\end{equation} 
behaves as $M \propto (\overline{k}-\overline{k}_c) \propto (q_c-q)$ near the condensation point (see Fig.~\ref{f5}). One can see that all the edges are in the condensate in the limit of $\overline{k}\to\infty$. (Note, however, that we consider a sparse network.)  

One can easily understand this condensation phenomenon. For the evolution of the number of edges in the condensate, $K_h(t)=Mt$, that is, the ``macroscopic'' number of edges attached to the hub, one can immediately write the following equation: 

\begin{equation}
\frac{dK_h}{dt} = \frac{K_h}{K}\left[\frac{K-K_h}{(q-1)t} - 2 \right]
\, .  
\label{e5a}
\end{equation} 
Here, $K(t)$ is the total degree of the network, and the second factor on the right hand side of the equation is simply the mean degree $\overline{k}_\text{n}$ of ``normal'' vertices 
(i.e., the hub is excluded) minus $2$. Indeed, according to rule (2) of the model, the probability that the hub will be chosen for the merging is $K_h/K$. Each act of merging, in average, increases the number of connections of the hub by $\overline{k}_\text{n}-2$, which explains the form of Eq.~(\ref{e5a}). 
Consequently, 

\begin{equation}
M = \frac{M}{2(qm-1)}\left[\frac{2(qm-1)-M}{q-1} - 2\right]
\, .  
\label{e5a}
\end{equation} 
One can see that this equation has a non-zero solution $M$ [exactly of the form (\ref{f5})] only if the growth rate parameter $q$ is less than the critical value $q_c$ given by expression (\ref{f4}). Note that similar equations may be written for 
for networks with splitting vertices (see below).  


\begin{figure}
\epsfxsize=67mm
\epsffile{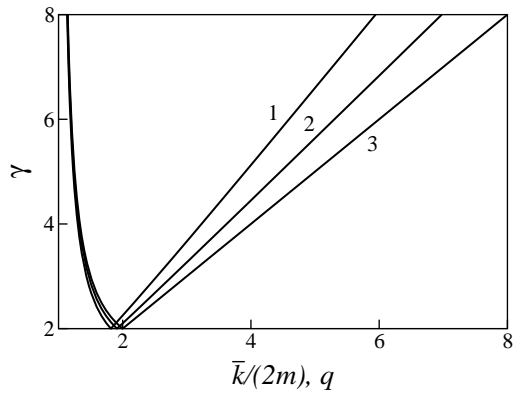}
\caption{
The $\gamma$ exponent of the degree distribution of network C as a function of $\overline{k}/(2m)$ and $q$. Curves 1, 2, and 3 correspond to $m=3,$ 6, and $\infty$, respectively. Note that the dependences on $\overline{k}/(2m)$ and $q$ are identical, but the ranges of the normal and the condensation phases are inversed [see expressions (\protect\ref{e6}) and (\protect\ref{e7})]. The condensation takes place for $\overline{k}/(2m) > \overline{k_c}/(2m) = 1 + \sqrt{1-1/m}$ and 
$q < q_c = 1 + \sqrt{1-1/m}$. 
}
\label{f6}
\end{figure}


``Normal vertices'' (i.e., with ``microscopic'' numbers of connections) have stationary power-law degree distributions (asymptotics) both in the normal and in the condensed phases. The $\gamma$ exponent of the degree distribution $P(k) \sim k^{-\gamma}$ is 

\begin{equation}
\gamma = 
2 + \frac{4\overline{k}m - \overline{k}^2 - 4m}{(\overline{k}-2)(\overline{k}-2m)} = 
2 + \frac{mq^2 - 2qm + 1}{q(m-1)} 
> 2
\label{e6}
\end{equation} 
in the phase without condensate ($\overline{k}<\overline{k}_c$, i.e. $q>q_c$) and 

\begin{equation}
\gamma = 
2 + \frac{\overline{k}^2 + 4m - 4\overline{k}m}{2\overline{k}(m-1)} = 
2 + \frac{2qm - mq^2 - 1}{(mq-1)(q-1)} 
> 2
\label{e7}
\end{equation} 
in the condensation phase,  
where $\overline{k}>\overline{k}_c$ ($q<q_c$). Fig.~\ref{f6} shows how the exponent of the degree distribution varies with the mean degree. 
Note that in both phases, near the condensation point, $\gamma-2 \propto |\overline{k}-\overline{k}_c| \propto |q-q_c|$. 

In the critical point, the degree distribution has the form: 

\begin{equation}
P(k) \sim \frac{1}{k^2\ln^2 k}
\, . 
\label{e8}
\end{equation} 
 
Thus, this network is scale-free both is the normal and in the condensation phases. The power-law form of the degree distribution in the condensation phase is rather unexpected. 

Let us explain this remark in more detail. 
A close analogy of the problem under consideration is the emergence of the giant connected component in a growing network, where a phase transition with the Berezinskii-Kosterlitz-Thouless singularity takes place   
\cite{chkns01,dms01,kkkr02,bb03}. In that case, the evolution equation for the size distribution of connected components is very similar to the evolution equation for the degree distribution in our case (see the next section). In this analogy, the giant connected component is analogous to the condensate of edges attached to a vertex, and the size distribution of finite connected components is analogous to the degree distribution in our case. The point is that the size distribution of connected components was found to be rapidly decreasing in the phase with the giant component (see Ref.~\cite{dms01}), while in contrast, in the present situation, the degree distribution is scale-free in the phase with the condensate.  

The paper \cite{ktms04} was mainly devoted to the simulation of the ``static'' version of quite similar network without multiple connections and loops of length $1$, $N=\text{const}$ with a growing number of connections, the sparse network regime. Scale-free degree distributions with exponents exceeding $2$, without any condensation, were reported. 
We do not consider precisely this situation here, since we focus on stationary degree distributions. These take place in the growing network.

\subsection{Network D}\label{ss-d}

At each time step (see Fig.~\ref{f7}): 


\begin{figure}
\epsfxsize=58mm
\epsffile{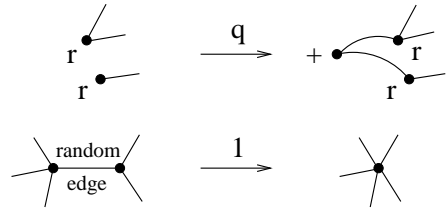}
\caption{
Processes creating network D. $m=2$.  
}
\label{f7}
\end{figure}


\begin{itemize}

\item[(1)]
$q>1$ new vertices are added to the network. Each of these vertices is attached to a randomly chosen vertex by $m$ edges.   

\item[(2)]
Simultaneously, the end vertices of a randomly chosen edge merge together, and this edge disappears.   

\end{itemize} 

At first sight, this model is close to model~C. The numbers of vertices and connections grow in the same way, and the mean degree is the same, Eq.~(\ref{e2}). The only difference is the way in which the merging vertices are chosen. 
One can treat this merging process as transformation of random edges with their end vertices into single vertices. 

In fact, there is an essential difference since now the choice of vertices, in principle, depends on correlations between the degrees of the nearest neighbor vertices in the network. 
Let us compare models C and D once again: (i) The evolution of model C produces degree-degree correlations but is not governed directly by them. (ii) The evolution of model D depends on the degree--degree correlations and, in its turn, produces these correlations.   

Strict calculations taking into account degree--degree correlations should be rather cumbersome. Furthermore, in related networks, studied in Ref.~\cite{ktms04}, correlations have not been found. So, we applied a simplifying ansatz: we assumed that correlations may be neglected. 


\begin{figure}
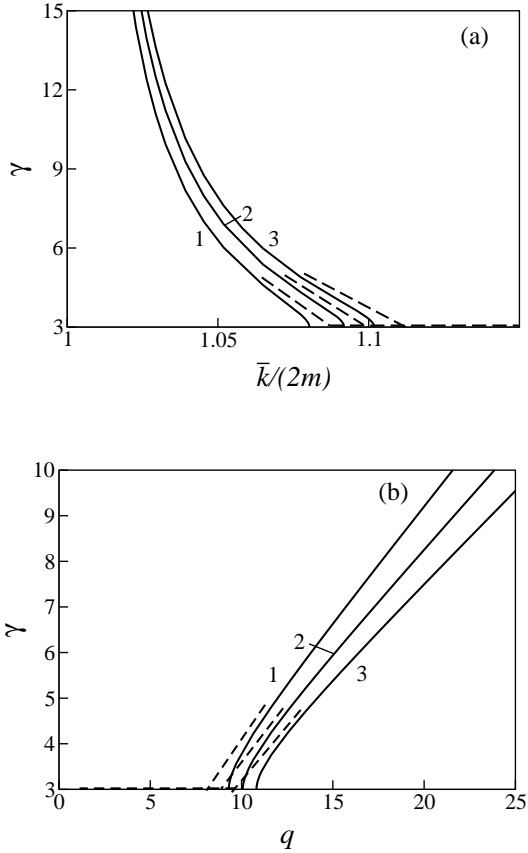

\epsfxsize=68mm
\epsffile{aggregation_fig8a.eps}
\epsfxsize=69.5mm
\epsffile{aggregation_fig8b.eps}
\caption{
The $\gamma$ exponent of the degree distribution of net\-work D as a function of $\overline{k}/(2m)$ (a) and $q$ (b). Curves 1, 2, and 3 correspond to $m=3,$ 6, and $\infty$, respectively. The solid lines are obtained by neglecting degree--degree correlations. The dashed lines qualitatively show results of the simulation. In the condensation phase, exponent $\gamma$ does not change with $\overline{k}$ and $q$, and is close to $3$. 
}
\label{f8}
\end{figure}


It turns out that with this assumption, equations for the degree distribution have a reasonable solution only if the mean degree of the network is below some value: 

\begin{equation}
\overline{k} < \overline{k}_c = 2.204m - 0.1115 + O(1/m) 
\, ,  
\label{e9}
\end{equation} 
or, equivalently, $q>q_c \cong 10.815-4.446/m$. In this region, our calculations provide a power-law degree distribution. The $\gamma$ exponent of this distribution approaches infinity at the minimal possible mean degree $2m$ (i.e., $q \to \infty$) and near $\overline{k}_c$ behaves as 

\begin{equation}
\gamma - 3 \sim \sqrt{\overline{k}_c - \overline{k}} \sim \sqrt{q-q_c}
\,   
\label{e10}
\end{equation} 
(see Fig.~\ref{f8}). This is in sharp contrast to the behavior $\gamma(\overline{k})$ near the critical point of network C (see Fig.~\ref{f6}). Note the value $3$ of exponent $\gamma$ at $k=\overline{k}_c$. 





\begin{figure}
\epsfxsize=86mm
\epsffile{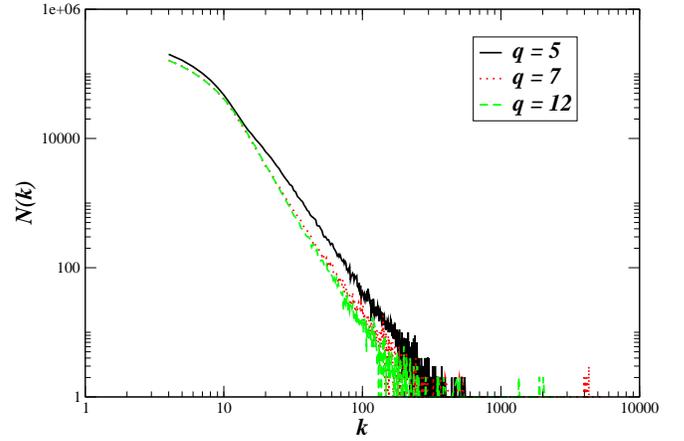}
\caption{
The average number of vertices of degree $k$ in simulated network D 
for $q=5,7,12$ and $m=3$.  
$P(k) = \overline{N}(k)/N$. 
The values $q=5$ and $q=7$ imply the presence of the condensate. 
The simulated networks contain up to $10^4$ vertices. 
} 
\label{f10}
\end{figure}

\begin{figure}
\epsfxsize=86mm
\epsffile{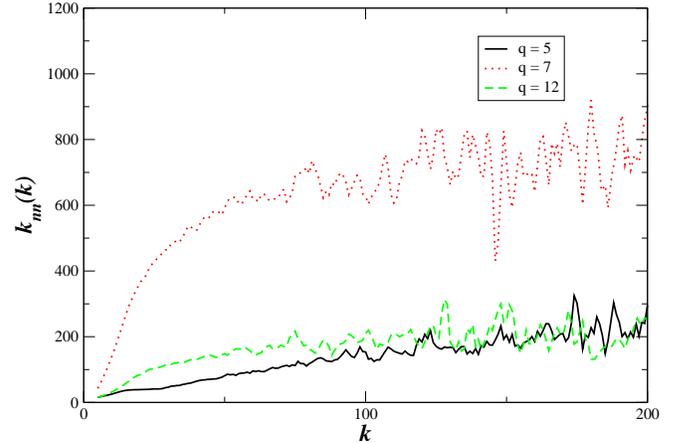}
\caption{
The dependence of the mean degree of the nearest neighbors of a vertex on the degree $k$ of this vertex, $\overline{k}_{nn}(k)$ in simulated network D 
for $q=5,7,12$ and $m=3$. 
The simulated networks contain up to $10^4$ vertices. 
}
\label{f10b}
\end{figure}


Above $\overline{k}_c$, with our assumption that the correlations are absent, the only solution was found to be pathological. This proves the importance of the correlations in this network, which may be especially important in the condensation phase.  

For studying these degree--degree correlations, we resort to numerical 
simulations following the rules of model D. The degree distributions of the resulting networks are shown in Fig.~\ref{f10}. Power-law-like degree distributions were observed both in the condensation phase and in the normal one. Fitting has given values of the $\gamma$ exponent slightly above $3$ at the studied values of the parameter $q$.   
We obtained the dependence of the mean degree of the nearest neighbors of a vertex on the degree $k$ of this vertex, $\overline{k}_{nn}(k)$.  
Of particular note here is that $\overline{k}_{nn}(k)$ 
has to be computed with care: the loops of length 1 are not to be considered for the vertex whose
nearest neighbors are under study. 
However, we take into account the nearest neighbors with one loops. This may be especially important in the condensation phase. The obtained dependences $\overline{k}_{nn}(k)$ are shown in Fig.~\ref{f10b}. The main conclusions are as follows:  

\begin{itemize}

\item[(1)] 
the network is correlated, and the degree--degree correlations are strong; 
the correlations are of assortative type; one reason for this is the
fact that for nodes with a high $k$ the self-loops contribute strongly
to the degree emphasizing such tendencies.

\item[(2)] 
the correlations are present both in the phase with the condensate and without it, Fig.~\ref{f10b} demonstrates a non-monotonous dependence of $\overline{k}_{nn}(k)$ on $q$; 

\item[(3)] 
the condensation phase transition takes place at higher values of the mean degree than $\overline{k}_c$, given by Eq.~(\ref{e10}); 

\item[(4)] 
the observed values of the $\gamma$ exponent of the degree distribution are restricted from above.  


 
\end{itemize} 


\subsection{Splitting vertices}\label{ss-split}


\begin{figure}
\epsfxsize=50mm
\epsffile{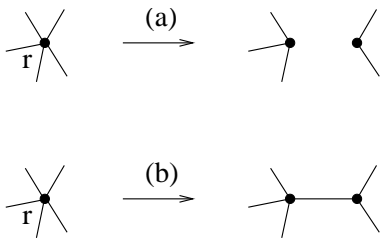}
\caption{
Splitting processes (examples): (a) splitting of a randomly chosen vertex into a pair of unconnected vertices; (b) splitting of a random vertex into a pair of connected vertices. 
In the simplest situation, each of possible resulting configurations occurs with equal probability. 
}
\label{f11}
\end{figure}


Instead of the process of the random attachment of new vertices in models A--C, one can introduce another channel of evolution---splitting (fragmentation) of vertices. In this paper we only touch upon two possibilities (see Fig.~\ref{f11}): 

\begin{itemize}

\item[(1)]
A randomly chosen vertex of degree $k$ splits into a pair of unconnected vertices of degrees $k'+k''=k$ in such a way that all possible resulting configurations are realized with equal probabilities.       

\item[(2)] 
A randomly chosen vertex of degree $k$ splits into a pair of vertices of degrees $k'+k''=k+2$, connected by an edge. Again, we assume that all possible resulting configurations occur with equal probabilities. 

\end{itemize} 
The number of these splittings per time step, in principle, may be not equal to $1$. 






\section{Derivations}\label{s-derivations}

In this section we describe details of our analytical calculations. 
We use an analytical technique similar to that for aggregation processes in more traditional systems \cite{bk95,kb00} (for various aspects of the aggregation processes, see, e.g., Refs.~\cite{t89,rdcb02,rm01,ha02}).

\subsection{Network A}\label{ss-a_d}

The evolution equation for the average number $\overline{N}(k,t)$ of vertices of degree $k$ in the network A at time $t$ has the following form: 

\begin{eqnarray} 
&&
\!\!\!\!\!\!\!
\overline{N}(k,t+1)  = 
\overline{N}(k,t) + \delta_{k,1} 
+ 
\frac{1}{N}\overline{N}(k-1,t) - \frac{1}{N}\overline{N}(k,t)
\nonumber
\\[5pt]
&& 
+ 
\left(\frac{1}{N}\right)^2 \sum_{k'+k''=k}\overline{N}(k',t)\overline{N}(k'',t) - \frac{2}{N}\overline{N}(k,t) 
\, .
\label{e11}
\end{eqnarray}  
Here the average is over the statistical ensemble. (A random network is a statistical ensemble: a set of configurations with their statistical weights.) 
Three first terms on the right-hand side of Eq.~(\ref{e11}) describe the process of the addition of a new vertex and the attachment it to a randomly chosen vertex. The two last terms describe the merging of a pair of randomly selected vertices. The factor $2$ in the last term is due to the fact that two vertices merge together. 

Note that we assume that our networks are large. So, a merging pair almost surely has no common nearest neighbors, and we can ignore the emergence of multiple connections during merging, if we do not interested in the cutoff region of the degree distribution.    

The total number of vertices in the network, $N$, is constant, so the evolution equation for the degree distribution $P(k,t) = \overline{N}(k,t)/N$ has the form: 

\begin{eqnarray} 
&&
\frac{\partial P(k,t)}{\partial t} = \delta_{k,1} 
+ 
P(k-1,t) - P(k,t)
\nonumber
\\[5pt]
&& 
+ 
\sum_{k'+k''=k}P(k',t)P(k'',t) - 2P(k,t) 
\, .
\label{e12}
\end{eqnarray} 
(as is usual, asymptotically long times are considered). Assuming a stationary form of the degree distribution (except of a time-dependent cutoff and of, maybe, a low-degree part of the distribution), we arrive at a stationary equation. 
The $Z$-transform of the degree distribution (a generating function) is 

\begin{equation}
n(z) \equiv  
\sum_{k=0} z^k P(k)
\, .
\label{e13}
\end{equation} 
So, in a $Z$-transformed form, we have 

\begin{equation}
0 = n^2(z) + (z-3)n(z) + z 
\, .
\label{e14}
\end{equation} 
The solution of this equation is 

\begin{equation}
n(z) = \frac{1}{2} [3-z - \sqrt{(9-z)(1-z)}] 
\, .
\label{e15}
\end{equation} 
This gives $n(1)=1$, as it should be with $\sum_k P(k)=1$. 
This root of the equation is chosen, since it must be $0<n'(0)=P(1)<1$. 
The equation gives $n'(0)=P(1)=1/3$. 

Near $z=1$, 

\begin{equation}
n(z) \cong \text{analytical terms}+(1-z)^{3/2-1}  
\, .
\label{e16}
\end{equation} 
This corresponds to the degree distribution $P(k) \sim k^{-3/2}$, Eq.~(\ref{e1}), since the form of a $Z$-transform near $z=1$ and the asymptotics of its original are related to each other in the following way  

\begin{eqnarray}
n(z\sim 1) \!\!\!\!\!&\cong&\!\!\!\!\! \text{analytical terms}+(1-z)^{\gamma-1} 
\nonumber
\\[5pt] 
&\longleftrightarrow& \ 
P(k \gg 1) \sim k^{-\gamma} 
\, , 
\label{e17}
\end{eqnarray} 
if $\gamma$ is non-integer. The result is rather typical not only for aggregation processes but also for general non-equilibrium networks, where the mean degree linearly grows with time (see discussion in Ref.~\cite{dmbook03}). 

Initially, we have assumed that the resulting degree distribution is stationary. 
In principle, one can made a more general assumption, e.g., for brevity, $P(k,t) = t^a f(k)$, where $a$ is some exponent, and $f(k)$ is an arbitrary function of $k$. 
After the substitution of this form into Eq.~(\ref{e12}) and $Z$-transformation, we find that the solution, $n(z,t)$, depends only on $z$, that is stationary.

\subsection{Network B}\label{ss-b_d}

The evolution equation for the mean number of vertices of degree $k$ in network B is 

\begin{eqnarray} 
&&
\overline{N}(k,t+1)  = 
\overline{N}(k,t) + q\delta_{k,m} 
\nonumber
\\[5pt]
&& 
+ 
\frac{qm}{N(t)}\overline{N}(k-1,t) - \frac{qm}{N(t)}\overline{N}(k,t)
\nonumber 
\\[5pt]
&& 
\!\!\!\!\!\!\!\!\!\!\!\!\!\!\!\!+ 
\frac{1}{N^2(t)} \sum_{k'+k''=k}\overline{N}(k',t)\overline{N}(k'',t) - \frac{2}{N(t)}\overline{N}(k,t)
\, .
\label{e18}
\end{eqnarray}
Here, the number of vertices $N(t) \cong (q-1)t$. The evolution equation for the degree distribution 
$P(k,t) \cong \overline{N}(k,t)/[(q-1)t]$ is 

\begin{eqnarray} 
&&
\!\!\!\!\!\!\!\!\!\!\!\!\!(q-1)[t\partial_tP(k,t) + P(k,t)] = q\delta_{k,m} 
+ qmP(k-1,t)  
\nonumber
\\[5pt]
&& 
\!\!\!\!\!\!\!\!\!\!\!\!\!- qmP(k,t)+ \sum_{k'+k''=k}P(k',t)P(k'',t) - 2P(k,t)
\, .
\label{e19}
\end{eqnarray}  
Assuming a stationary degree distribution, in a $Z$-transformed form, this is 

\begin{equation}
0 = q z^m + n^2(z) + qmz n(z) - [q(m+1)+1]n(z) 
\, 
\label{e20}
\end{equation} 
[compare with Eq.~(\ref{e14}) for $q=1$]. The solution of Eq.~(\ref{e20}) is analytical at $z=1$, so that the degree distribution is a rapidly decreasing function. Suppose, however, that the rate $q$ is close to $1$. Then deviations from non-analytical behavior (\ref{e16}) of $n(z)$ are observed only in the range $1-z \lesssim (q-1)^2/m$. This results in a size-independent cutoff $k_{\text{cut}} \sim m/(q-1)^2$ of the power-law dependence $P(k) \sim k^{-3/2}$.

\subsection{Network C}\label{ss-c_d}

The evolution equation for the mean number of vertices of degree $k$ in this network is of the form:  

\begin{eqnarray} 
&&
\overline{N}(k,t+1)  = 
\overline{N}(k,t) + q\delta_{k,m} 
\nonumber
\\[5pt]
&& 
+ 
\frac{qm}{N(t)}\overline{N}(k-1,t) - \frac{qm}{N(t)}\overline{N}(k,t)
\\[5pt]
&& 
+ \frac{1}{N^2(t)\overline{k}} \sum_{k'+k''=k+2}\overline{N}(k',t)k''\overline{N}(k'',t) 
\nonumber
\\[5pt]
&& 
- \frac{1}{N(t)}\overline{N}(k,t) - \frac{k}{N(t)\overline{k}}\overline{N}(k,t)
\nonumber
\, .
\label{e21}
\end{eqnarray} 
Note that while deriving this equation, we did not assume the absence of correlations between the nearest neighbor vertices. Rule (2) of the model ensures that the second vertex for merging (a random nearest neighbor of a random vertex) is chosen with the probability proportional to the degree of this vertex. This is true irrespective of the presence of correlations in the network. 
 
For the stationary degree distribution, in a $Z$-tran\-s\-formed form, we have 

\begin{eqnarray}
&&
0 = q z^m + \frac{1}{z^2\overline{k}}n(z)zn'(z) - \frac{1}{\overline{k}}zn'(z)
\nonumber
\\[5pt]
&& 
+ qmz n(z) - q(m+1)n(z) 
\, .
\label{e22}
\end{eqnarray} 
[Note that the $Z$-transform of the convolution with $\sum_{k'+k''=k+2}P(k')k''P(k'')$ is 
$$
\frac{1}{z^2}\{n(z)zn'(z) - P(0)0P(0) -z(P(0)1P(1)+P(1)0P(0))\} 
\, .
$$ 
We have $P(0)=0$, so that this expression simplifies.]  
Recall that 

\begin{equation}
q = \frac{\overline{k}-2}{\overline{k}-2m} > 1
\, .
\label{e23}
\end{equation} 
Equation~(\ref{e22}) is the Abel equation of the second kind. In a canonical form it looks as 

\begin{equation}
[z^2-n(z)]n'(z) = 
\overline{k}q [z^{m+1} + (mz^2 - (m+1)z)n(z)] 
\, .
\label{e24}
\end{equation} 
For the detailed analysis of a very similar equation for the size distribution of finite connected components in growing networks, see Ref.~\cite{dms01}. Note, however, that in general terms, Eq.~(\ref{e24}) has a different solution than that in Ref.~\cite{dms01}.  

If $P(k=0)=0$, Eq.~(\ref{e24}) implies 

\begin{equation}
n(z\sim0) \cong 
\frac{\overline{k}(\overline{k}-2)}
{\overline{k}^2(m+1)-\overline{k}(2m+1)-2m^2}
z^m
\, .
\label{e25}
\end{equation} 

For obtaining a large $k$ asymptotics of $P(k)$, 
one must find the solution $n(z)$ of Eq.~(\ref{e24}) with the initial condition (\ref{e25}) near $z=0$. One can check that this solution arrives at $1$ at $z=1$: $n(1)=1$. 

The linearization of Eq.~(\ref{e24}) near $z=1$ shows that $n(z)$ linearly approaches $z=1$, with a derivative $n'(1)$,  
which satisfies the equation:
\begin{equation}
[-2 + n'(1)]n'(1) = \frac{\overline{k}(\overline{k}-2)}{\overline{k}-2m}[-2m + n'(1)]
\, .
\label{e26}
\end{equation} 
The meaning of $n'(1)$ is the mean degree of a vertex with degree $k=o(N)$, that is the mean degree of a ``normal'' vertex.  
The solution of this square equation is 
\begin{equation}
n'(1) = \frac{\overline{k}^2 - 4m - |-\overline{k}^2 + 4\overline{k}m - 4m|}{2(\overline{k}-2m)}
\, .
\label{e27}
\end{equation} 
That is, there is a point $\overline{k}_c$ (and the corresponding $q_c$), satisfying the equation 
\begin{equation}
\overline{k}_c^2 - 4\overline{k}_cm + 4m=0
\, , 
\label{e28}
\end{equation}  
where the expression for $n'(1)$ changes its form. $\overline{k}_c$ and $q_c$ 
are given by Eqs.~(\ref{e3}) and (\ref{e4}), respectively. 

\begin{figure}
\epsfxsize=48mm
\epsffile{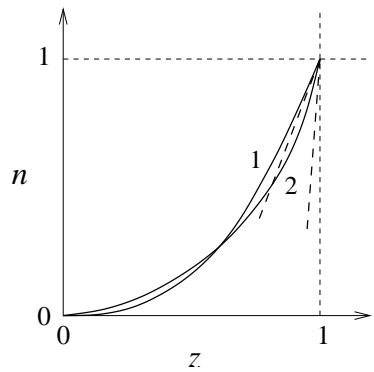}
\caption{
The schematic view of the solution of Eq.~(\protect\ref{e24}) in the normal (1) and condensation (2) phases. The dashed lines show the corresponding $1-\overline{k}(1-z)$ dependences. In the condensation phase, the slope of the solution at $z=1$ is lower than $\overline{k}$. 
}
\label{f14}
\end{figure}

At $\overline{k}<\overline{k}_c$, the slope $n'(1)$ equals the mean degree of the network, $\overline{k}$. That is, all the vertices of the network are ``normal''. Above $\overline{k}_c$ (i.e., at $q<q_c$) $n'(1) = 2qm = 2m(\overline{k}-2)/(\overline{k}-2m) < \overline{k}$. This means that a fraction $M = [\overline{k}-n'(1)]/\overline{k}$ [see expression (\ref{e5})] is in the condensed state. 
The variation of the solution of Eq.~(\ref{e24}) with $\overline{k}$ is shown schematically in Fig.~\ref{f14}. 


The large-degree asymptotic behavior of the degree distribution of ``normal vertices'' is obtained by analyzing the form of the solution of Eq.~(\ref{e24}) near $z=1$. 
We pass to new variables $z = 1-\xi$, $n(z) = 1 - n'(1)\xi + v(\xi)$ in Eq.~(\ref{e24}). The inspection of the resulting equation shows that at $\overline{k} \neq \overline{k}_c$, the contribution $v(\xi)$ is a power-law function at small $\xi$, $v(\xi) \propto \xi^a$, where exponent $a$ is greater than $1$. 
The substitution of this form into the equation allows us to obtain $a$. Using the correspondence (\ref{e17}) results in the formulas (\ref{e6}) and (\ref{e7}) for the $\gamma$ exponent of the degree distribution in the normal and condensed phases, respectively (see Fig.~\ref{f6}). 

In the critical point, $\overline{k} = \overline{k}_c$, the solution of the equation for $v(\xi)$ has a more complex form with an additional logarithmic factor (for more detail, see Ref.~\cite{dms01}). 
The resulting solution $n(z\sim1)$ is  

\begin{equation}
n(z) \cong 1 - \overline{k}_c (1-z) + 
\text{const}\,\frac{1-z}{\ln[\text{const}(1-z)]}
\, . 
\label{e29}
\end{equation}  
The asymptotics of the original of this $Z$-transform is

\begin{eqnarray}
&&
P(k) = \oint_c \frac{dz}{2\pi i}z^{-k-1} P(k) 
\nonumber
\\[5pt]
&& 
\propto  
\oint_c \frac{dz}{2\pi i} \frac{1-z}{\ln[\text{const}(1-z)]}z^{-k-1} 
\cong 
\oint_{c'} \frac{ds}{2\pi i} \frac{s}{\ln s} e^{sk}
\nonumber
\\[5pt]
&&  
= 
\int_0^{\infty} \frac{ds}{2\pi i}\,s \left(\frac{1}{\ln s} - \frac{1}{\ln s + 2\pi i}\right)e^{-sk}
\nonumber
\\[5pt]
&& 
\cong \int_0^{\infty} \frac{ds}{2\pi i}\,s \frac{2\pi i}{\ln^2 s}e^{-sk} 
\sim \frac{1}{k^2\ln^2 k}
\, . 
\label{e30}
\end{eqnarray} 
Here the 
contour $c$ is around $0$, within the unit circle.    
The contour $c$ is deformed to the contour $c'$, which comes from $-\infty$ to $+0$ along the cut of $\ln$ and then returns to $-\infty$ by the other shore of the cut.  
This deformation was made to ensure the decrease of the exponent. 
Thus, we have arrived at the degree distribution (\ref{e8}) at the critical point. 

One can show that at $\overline{k} \neq \overline{k}_c$, in the large networks, the degree distribution has two regions. 
(i) The power-law dependence with exponents (\ref{e6}) and (\ref{e7}) is realized in the range of degrees 
$\ln k > 2(\overline{k}^2 - 4\overline{k} + 4m)/|\overline{k}^2 - 4\overline{k}m + 4m|$. (ii) At smaller degrees, the critical dependence (\ref{e8}) is present.

\subsection{Network D}\label{ss-d_d}

Merging vertices in this model are the ends of randomly selected edges. 
In this case, for the strict description of the network evolution one has to solve an equation for a joint distribution $P(k',k'')$ of the degrees of the nearest neighbor vertices. This is an essentially more hard problem than the analysis of Eq.~(\ref{e12}), (\ref{e19}), and (\ref{e21}) for $P(k)$ or $\overline{N}(k)$ in previous sections. Instead of making these cumbersome calculations, we ignore the possibility of the degree--degree correlations and check whether this ansatz leads to reasonable results or not. This simplification allows us to consider a more simple evolution equation. 

Assuming the absence of correlations between degrees of the nearest neighbors of network D results in the following rate equation:  

\begin{eqnarray} 
&&
\overline{N}(k,t+1)  = 
\overline{N}(k,t) + q\delta_{k,m} 
\nonumber
\\[5pt]
&& 
+ 
\frac{qm}{N(t)}\overline{N}(k-1,t) - \frac{qm}{N(t)}\overline{N}(k,t)
\nonumber
\\[5pt]
&& 
+ \frac{1}{N^2(t)\overline{k}^2} \sum_{k'+k''=k+2}k'\overline{N}(k',t)k''\overline{N}(k'',t) 
\nonumber
\\[5pt]
&& 
- 2\frac{k}{N(t)\overline{k}}\overline{N}(k,t)
\, .
\label{e31}
\end{eqnarray} 
For brevity, we consider only the case $m>2$ (some of the resulting formulas may be different in the cases of $m=1,2$). 
So, for a stationary degree distribution, we have in a $Z$-transformed form: 

\begin{eqnarray}
&&
0 = q z^m + \frac{1}{z^2\overline{k}^2}[zn'(z)]^2 
- \frac{2}{\overline{k}}zn'(z) 
\nonumber
\\[5pt]
&& 
+ qmz n(z) - [q(m+1)-1]n(z) 
\, .
\label{e32}
\end{eqnarray} 
This is a nonlinear differential equation of the first order. It crucially differs from the corresponding Eq.~(\ref{e22}) for the model C, and provides a different set of behaviors. 

Near $z=0$, the solution of Eq.~(\ref{e32}) is of the following form: 

\begin{equation}
n(z \sim 0) \cong 
\frac{\overline{k}(\overline{k}-2)}{2\overline{k}(m-1) - 4m^2 + \overline{k}^2 m} z^m 
\, . 
\label{e33}
\end{equation} 
This asymptotics is used as a boundary condition for Eq.~(\ref{e32}). 


\begin{figure}
\epsfxsize=48mm
\epsffile{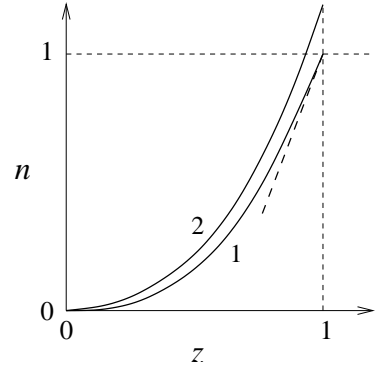}
\caption{
The solution of Eq.~(\protect\ref{e32}) below (1) and above (2) the ``critical'' point $\overline{k}_c$. Only for $\overline{k}<\overline{k}_c$ (i.e., $q>q_c$), the solution provides $n(1)=1$ and the limit slope $\overline{k}$ (see the dashed line). 
}
\label{f15}
\end{figure}


For a numerical analysis, the following form of Eq.~(\ref{e32}) is more convenient: 

\begin{eqnarray}
&&
n'(z) = kz - k\left\{z^2 - \frac{1}{\overline{k}-2m}[\overline{k}^2(\overline{k}-2)z^m \right.
\nonumber
\\[5pt]
&& 
\left. 
\phantom{\frac{1}{|}}+ ((\overline{k}-2)mz-\overline{k}+2m)n(z)] \right\}^{1/2}
\, , 
\label{e34}
\end{eqnarray} 
Note that only the solution with `minus' is reasonable. 
One can check that at the boundary $z=1$, if $n(1)=1$, then Eq.~(\ref{e32}) gives $n'(1)=\overline{k}$ and vice versa, if $n'(1)=\overline{k}$, then Eq.~(\ref{e32}) gives $n(1)=1$. 
On the other hand, the numerical solution of Eq.~(\ref{e34}) shows that the solution with $n(1)=1$ exists only at sufficiently low $\overline{k}$. 
Above some value $\overline{k}_c$, the solutions of Eq.~(\ref{e32}) or Eq.~(\ref{e34}) turn out to be greater than $1$ at some $z$: $n(1)>1$ (see Fig.~\ref{f15}, compare with Fig.~\ref{f14}). 
But this is impossible, since surely $\sum_k P(k) = 1$. 
This contradiction indicates that our assumption does not hold at least for $\overline{k}>\overline{k}_c$. 

Let us study the analytical structure of the solution $n(z)$, $z$ near $1$, for $\overline{k}<\overline{k}_c$. As in Sec.~\ref{ss-c_d}, introducing new variables, $z=1-\xi$, $n(z) = 1 - \overline{k}\xi + v(\xi)$, we pass to an equation for $v(\xi)$. It turns out, however, that unlike Sec.~\ref{ss-c_d}, this equation has no solution $v(\xi) \sim \xi^a$ with exponent $1<a<2$. So, we have to search for the solution in the analytical form $v(\xi)=C\xi^2$. Substituting this form into the equation gives 

\begin{eqnarray}
&&
C = \frac{\overline{k}(\overline{k}+\overline{k}m-4m)}{4(\overline{k}-2m)}
\nonumber
\\[5pt]
&& 
\!\!\!\!\!\!\!\!\!\!\!\!\!
\times 
\left[
1 - \sqrt{1 - 2\frac{(\overline{k}-2)(\overline{k}-2m)m(2\overline{k}+m-1)}{(\overline{k}+\overline{k}m-4m)^2}\,}
\right]
\!. 
\label{e35} 
\end{eqnarray} 
(One can check that only the sign ``minus'' at the square root leads to reasonable final values of the exponent of the degree distribution.) $C$ is real only if $\overline{k} \leq \overline{k}_c$, where $\overline{k}_c$ is the point where the square root in Eq.~(\ref{e35}) is zero. $\overline{k}_c$ is exactly the point above which the solution $n(z)$ is not reasonable [i.e., where $n(1)>1$]. 
The resulting expression for $\overline{k}_c$ is cumbersome but for $m \gg 1$  
we have

\begin{equation}
\overline{k}_c \cong \frac{7+\sqrt{113}}{8}m - \frac{3-11/\sqrt{113}}{7+\sqrt{113}} 
\, ,  
\label{e36}
\end{equation} 
which leads to the formula (\ref{e9}). 

So, if $\overline{k} \leq \overline{k}_c$, we can substitute $v(\xi)=C\xi^2+w(\xi)$ into the equation for $v(\xi)$. An inspection of the resulting equation for $w(\xi)$ shows that $w(\xi) \cong D\xi^b$, where exponent $b>2$. One can easily find $b$:  

\begin{equation}
b = \frac{\overline{k}^2(m-1)}{(\overline{k}-2m)(2C-\overline{k})}
\,   
\label{e36a}
\end{equation} 
with $C$ given by Eq.~(\ref{e35}), 
and by using the correspondence rule (\ref{e17}) we readily obtain the exponent of the degree distribution:   
\begin{eqnarray}
&&
\gamma = 3 +  
\nonumber
\\[7pt]
&& 
\!\!\!\!\!\!\!\!\!\!\!\!\!\!\!\!\!\!\!\!
\frac{2\sqrt{1 - 2\dfrac{(\overline{k}-2)(\overline{k}-2m)m(2\overline{k}+m-1)}{(\overline{k}\!+\!\overline{k}m\!-\!4m)^2}}}
{\dfrac{\overline{k}(m\!-\!1)}{\overline{k}\!+\!\overline{k}m\!-\!4m}\!-\!\sqrt{1\!-\!2\dfrac{(\overline{k}\!-\!2)(\overline{k}\!-\!2m)m(2\overline{k}\!+\!m\!-\!1)}{(\overline{k}\!+\!\overline{k}m\!-\!4m)^2}}}
\, . 
\label{e37}  
\end{eqnarray} 
This $\gamma(\overline{k})$ dependence is shown in Fig.~\ref{f8}. 

The analytical results for network D were obtained in the framework of the simplifying ansatz: we ignored degree--degree correlations. 
The simulation of this network has shown that the correlations are significant (see Sec.~\ref{ss-d}) and that our analytical predictions for the normal must be corrected. Furthermore, the simulations have allowed us to describe the structure of the network in the condensation phase.

\subsection{Splitting vertices}\label{ss-splitr_d}

Here we only show the contributions due to splitting processes to evolution equations for the degree distribution. Splitting process (1) of Sec.~\ref{ss-split} generates with equal probabilities all possible configurations of two unconnected vertices [see Fig.~\ref{f11} (a)].  
For example, the splitting of a vertex of degree $4$ produces: (i) with probability $1/8$, a pair of vertices of degrees $0$ and $4$, (ii) with probability $4/8$, a pair of vertices of degrees $1$ and $3$, and (iii) with probability $3/8$, a pair of vertices of degrees $2$ and $2$. One can easily write down the probability of resulting configurations for any degree of the splitting vertex. 
This allows us to find the probability that a vertex of degree $k$ will emerge due to the splitting of a vertex of degree $q$. Finally, instead, e.g., of the contribution $\delta_{k,1} + P(k-1,t)-P(k,t)$ due to the process of random attachment of a vertex, we have the following terms: 
\begin{equation}
\sum_{q=0}\frac{1}{2^{q-1}} {q \choose k}P(q,t) - P(k,t)
\, . 
\label{e38}
\end{equation}
Note that actually only terms $q \geq k$ are nonzero in this sum.
In a $Z$-transformed form, this is 
\begin{equation}
2 n\left(\frac{1+z}{2}\,,t\right) - n(z,t)
\, .
\label{a39}
\end{equation} 

In splitting process (2) of Sec.~\ref{ss-split}, an emerging pair of vertices is interconnected by an extra edge. This leads, instead of the contribution of the form (\ref{e38}), to the following terms: 
\begin{equation}
\sum_{q=0}\frac{1}{2^{q-1}} {q \choose k-1}P(q,t) - P(k,t)
\, ,  
\label{e40}
\end{equation}
where only terms $q \geq k-1$ are nonzero in this sum. 
In a $Z$-transformed form, Eq.~(\ref{e40}) looks as   
\begin{equation}
2 z\, n\left(\frac{1+z}{2}\,,t\right) - n(z,t)  
\, . 
\label{e41}
\end{equation}  

Formulas (\ref{e38})--(\ref{e41}) may be used to modify any of the evolution equations in the preceding sections. The resulting equations, in a $Z$-transformed form, will be non-local due to the $n((1+z)/2)$ term. We do not analyze these functional equations in the present paper. 

\section{Discussion and summary}\label{s-summary}

Several remarks are necessary. We have studied a representative set of models allowing an analytical solution
if no strong correlations are present.
For sake of simplicity and brevity, we considered only models, generating stationary (at least is some range of degrees) degree distributions.  
We have indicated that we had to ignore correlations in one of networks, and have demonstrated that this neglect created some problems. In addition, we ignored multiple connections and loops of length $1$, which, in principle, emerge during the evolution of these networks. 

The effect of multiple connections and loops of length $1$ in related network models was discussed in Ref.~\cite{bk03}. Several features of the network structure depend on the presence (or absence) of these configurations of edges. In particular, the position of a size-dependent cutoff of the degree distribution may change. The number of vertices, attracting the condensate of edges may also change. We did not considered these details (cutoffs, the precise number of vertices attracting the condensate, etc.). So, the strict accounting for multiple connections and one-loops should not essentially change the conclusions of this paper. 

One should emphasize the difference of the condensation transitions considered in this work from those which were observed earlier in networks.  
We have shown that in networks C and D, scale-free degree distributions are realized both in the normal and in the condensation phases. In contrast, in equilibrium networks, a scale-free degree distribution exists only in the condensation phase, and a degree distribution is rapidly decreasing in the normal phase \cite{dmbook03,zjj02}.  

The merging of vertices is only one of processes in the networks under discussion. This process reduces the number of vertices, so other processes, increasing this number, must be present. In this paper, as a parallel process, we used an injection of new vertices (with subsequent attachment to existing vertices) and fragmentation (splitting). This subject has been analyzed in the
aggregation literature \cite{vz88,stc92,stc94}.
Also, one can implement duplication (copying) or partial copying of vertices \cite{kkkr02,dms02,spsk02} or other processes. 
One should note, that correlations are important in low-dimensional
aggregation processes. In this respect we have an analogy with
this case. In the usual aggregation theory the analysis proceeds
by considering the scaling properties of the aggregation or fragmentation
kernels (see e.g. \cite{ha02} and references therein). For models such as 
those considered in our paper 
these are not known
a priori and in the presence of degree correlations arise
``self-consistently''.

As is natural, a uniform picture for all networks created by aggregation processes cannot be obtained in principle. However, the models that we have considered in the present paper, reveal a number of basic features:    

\begin{itemize}

\item[(1)]
The aggregation easily generates network architectures where hubs play a profound role. 
 
\item[(2)]
The aggregation often leads to gelation, or, in other words, to condensation of edges in these networks. We have found the condensation point at some mean degree value and have traced the variation of network structure in the normal and the condensation phases. 

\item[(3)]
These networks are evolving networks, and so their structure is characterized by strong correlations, in particular, by strong degree--degree correlations of assortative type.   

\end{itemize} 
These features, which we observed by using demonstrative models, should be present in more complex networks of this \vspace{-10pt}kind.

\begin{acknowledgments}
MJA is grateful to the Center of Excellence program
of the Academy of Finland for support. SD would like
to acknowledge the hospitality of the Helsinki University
of Technology. A part of the work was made when one of the authors (SD) attended the Exystence Thematic Institute on Networks and Risks (Collegium Budapest, June 2004). 
SD was partially supported by project POCTI/FAT/46241/2002 and  
POCTI/MAT/46176/2002. 
SD thanks Kim Sneppen for a useful discussion in Budapest,
and MJA thanks Supriya Krishnamurty for useful comments.
\end{acknowledgments}

\end{document}